\documentclass[
 reprint,
 amsmath,amssymb,
 aps,floatfix]{revtex4-2}
\usepackage{comment}%
\usepackage{graphicx}% Include figure files
\usepackage{dcolumn}
\usepackage{bm}

\usepackage{xcolor}

\usepackage[UKenglish]{babel}

\def\vec#1{{\ensuremath{\bm{#1}}}}
\def\n{\ensuremath{\hat{\bm{n}} }}
\def\nperp{\ensuremath{\hat{\bm{n}}^* }}

\def\half{{\textstyle \frac{1}{2}}}

\def\be{\begin{equation}}
\def\ee{\end{equation}}
\def\bea{\begin{eqnarray}}
\def\eea{\end{eqnarray}}

\def\nl{\hfil\break}

\begin{document}

\preprint{APS/123-QED}

\title{Metric mechanics with non-trivial topology: actuating irises, cylinders and evertors}

%\affiliation{Department of Engineering, University of Cambridge, Trumpington St, Cambridge CB2 1PZ, U.K.} 


\author{D.\ Duffy$^1$}
\author{M.\ Javed$^{2,3}$}
\author{M.\ K. \ Abdelrahman$^{2,4}$}
\author{T.\ H.\ Ware$^{2,3,4}$} % other authors and affiliations in email
\author{M.\ Warner$^5$}
\author{J.\ S.\ Biggins$^1$}

%\affiliation{
% This line break forced% with \\}%
%\affiliation{
% Third institution, the second for Charlie Author}%
\affiliation{$^1$Department of Engineering, University of Cambridge, Trumpington St., Cambridge CB2 1PZ, U.K.}
\affiliation{$^2$Department of Bioengineering, University of Texas at Dallas, Richardson, TX 75080, USA }
\affiliation{$^3$Department of Biomedical Engineering, Texas A\&M University, College Station, TX 77843, USA}
\affiliation{$^4$Department of Materials Science and Engineering, Texas A\&M University, College Station, TX 77843, USA}
\affiliation{$^5$Department of Physics, University of Cambridge, 19 JJ Thomson Ave., Cambridge CB3 0HE, U.K.}

\begin{abstract}
Liquid crystal elastomers contract along their director on heating and recover on cooling, offering great potential as actuators and artificial muscles. If a flat sheet is programmed with a spatially varying director pattern, it will actuate into a curved surface, allowing the material to act as a strong machine such as a grabber or lifter. Here we study the actuation of programmed annular sheets which, owing to their central hole, can sidestep constraints on area and orientation. We systematically catalogue the set of developable surfaces encodable via axisymmetric director patterns, and uncover several qualitatively new modes of actuation, including cylinders, irises, and everted surfaces in which the inner boundary becomes the outer boundary after actuation. We confirm our designs with a combination of experiments and numerics. Many of our actuators can re-attain their initial inner or outer radius upon completing actuation, making them particularly promising, as they can avoid potentially problematic stresses in their activated state even when fixed onto a frame or pipe. \vspace{\baselineskip}
\\
\centering{Accepted 22 Nov.~2021, published in Phys.~Rev.~E \textbf{104}, 065004 (2021)\\ \url{https://link.aps.org/doi/10.1103/PhysRevE.104.065004} \\  \textcopyright 2021 American Physical Society}
\end{abstract}

\maketitle

%\tableofcontents
\section{Introduction}
In classical engineering, a mechanism is a device for transforming motion/force from one form to another \cite{brownfive}; for example, a screw translates rotation to linear motion, and a lever trades displacement for force. Tables of ingenious mechanisms were drawn up in the nineteenth century (see \url{http://507movements.com/}) and a skilled mechanical engineer can combine these building blocks into a useful machine. In soft matter a similar challenge has emerged. We have a growing class of active shape-changing materials --- swelling hydrogels \cite{hirokawa1984volume}, contracting nematic elastomers \cite{warner2007liquid}, dilating dielectric elastomers \cite{pelrine2000high}, inflating pneumatics \cite{martinez2012elastomeric} --- and are challenged with transforming these elementary shape changes into alternative modes of actuation with desirable mechanical properties. How do you transform contraction into a push, or swelling into a squeeze? 

The key to such transformations is assembling complementary shape changes into an actuating mechanism. This approach is exemplified by Harrison's bimetallic strip, which turns differential expansion into bend \cite{timoshenko1925analysis} and thereby trades force for displacement, just like a lever. In soft materials, complementary shape changes can often be programmed into different regions of a single sample. For example, spatially patterning the  cross-link density in hydrogels gives patterned magnitudes of dilation on swelling/deswelling \cite{klein2007shaping, kim2012designing, gladman2016biomimetic}, while spatially patterning the alignment direction in  liquid crystal elastomers (LCEs) gives patterned directions of contraction on heating \cite{de2012engineering, ware2015voxelated,ambulo2017four}. These programmed materials can then achieve sophisticated modes of actuation: the material is the mechanism, if not quite yet the machine \cite{bhattacharya2005material}.

Here we focus on LCEs: rubbery networks of rod-like repeat units which spontaneously align along some direction $\n$ forming a nematic LC phase within the material \cite{warner2007liquid}. Following the protocol in \cite{boothby2017dual}, we fabricate LCE sheets by UV-crosslinking a nematic oligomer sandwiched between two sheets of glass. The inner faces of the glass sheet are photo-patterned with a preferred planar nematic alignment $\n(\vec{x})$, which is passed to the nematic fluid and crosslinked in as it forms the elastomer. After the elastomer sheet is released, this alignment can be reversibly disrupted with heat (reflecting the nematic-isotropic LC phase transition) causing the elastomer to contract by a factor of $\lambda \sim 0.65$ along $\n$. Contraction is accompanied by a transverse dilation $\lambda^{-\nu}$, with the opto-thermal Poisson ratio being strictly $\nu=1/2$ in the volume-preserving response characteristic of LCEs, although it can rise as high as $\nu=2$ in nematic photo-glasses \cite{liu2015new}, and can take a range of other values in pneumatic or swelling sheets\cite{warner2020inflationary, urayama_swelling}.

Several elementary entries in a table of LCE mechanisms are now well established. First are monodomains: sheets with uniform planar alignment which simply contract on heating, and can pull a load as they do so \cite{kupfer1991nematic}, failing however, because of Euler instability, to push as they elongate during recovery.  Second are benders, which arise whenever a thin strip or film suffers differential shape change through its thickness. In LCE sheets, thickness variation can be achieved by programming different director alignments on each side of the cross-linking cell \cite{modes2010anisotropic, van2007glassy, sawa2011shape}, or by the asymmetric actuation of a monodomain through its thickness \cite{white2008high}.  All these benders produce  high-displacement, low-force motion that looks impressive but, like the original bimetallic thermostats, is better suited to sensing than mechanical work. Third are pushers and lifters, created by programming a flat sheet with an alignment that varies in-plane and morphs the sheet into a curved-surface on activation. For example, a disk programmed with concentric circles of contraction (a topological defect with director winding number $m=1$) will rise into a cone on activation \cite{modes2011gaussian,de2012engineering,de2012engineering, ambulo2017four}. These surfaces have fundamentally different metrics to the original flat sheets \cite{klein2007shaping, modes2011gaussian}, allowing them to have Gauss (intrinsic) curvature that can only be flattened by energetically expensive stretch. Lifters mostly arise when the Gauss curvature is positive (cones, caps, spindles etc.) causing a protrusion on activation (though an evolving hyperbolic cone also protrudes, despite its negative GC away from the origin), and \emph{metric mechanics} \cite{warner2020topographic} makes such pushers mechanically strong: LCE cones can lift thousands of times their own weight as they rise \cite{guin2018layered}. A final category could be radial $m=1$ defect patterns, which  buckle into ruff-like ``anticones'' with negative Gauss curvature, although it is unclear what mechanical utility such actuation offers.

Several recent works have focused on the programming of complex surfaces and inverse design \cite{warner2018nematic, griniasty2019curved, fengbigginswarner}, demonstrating, for example, how to choose a pattern of contraction that morphs a sheet into a face \cite{aharoni2018universal}. Though analytically virtuosic, mechanically these are variations on the theme of cones and caps. Here we take a different approach, and investigate the elementary modes of actuation of annular sheets. The introduction of a hole fundamentally changes  the sheet's topology, allowing simple but qualitatively different modes of actuation, and inviting use cases such as apertures, sphincters, filters and pipes. Here we focus on axisymmetric  director patterns encoding Gauss-flat shapes. Any simply connected Gauss-flat surface can be flattened into the plane isometrically (i.e.~without stretch). However, introducing a hole opens a rich design space of actuated shapes, including flat irises, truncated cones/anticones, cylinders and everted annuli. Although $\nu$ is limited to 1/2 in our LCE experiments, our treatment shows that the same family of shapes arises for shape-changing materials with any $\nu$. Importantly, despite producing Gauss-flat surfaces, such annular actuations are strong, and cannot be blocked without energetically expensive stretch. 

\section{Shape programming of annular LCE actuators}

\begin{figure}[ht]
\centering
\includegraphics[width=\linewidth]{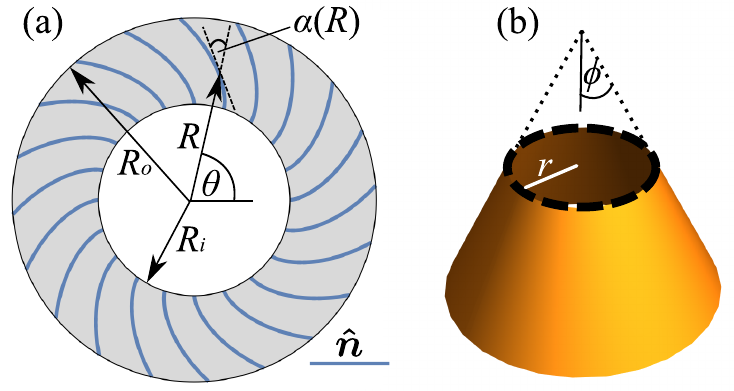}
  \caption{(a) An annular LCE sheet in the reference domain, with an axisymmetric spiral director pattern characterized by the angle $\alpha(R)$. (b) A truncated cone with semi-angle $\phi$. The dashed circle has curvature $1/r$, and geodesic curvature $k_g$ of magnitude $\sin(\phi)/r$.}
  \label{fig:theory}
\end{figure}

In this work, we consider an initially flat annular LCE sheet, $R_i<R<R_o$, patterned with a planar nematic director field $\n$, as seen in Fig.~\ref{fig:theory}a. Upon heating, the sheet contracts by $\lambda$ in the direction parallel to $\n$, and extends by $\lambda^{-\nu}$ in the orthogonal direction. An infinitesimal length element in the undeformed sheet, $\mathrm{d}\mathbf{l}$, thus has an activated length given by
\begin{equation}
\mathrm{d}l_A^2=\mathrm{d}\mathbf{l} \cdot  (\lambda^2 \, \n \otimes \n+\lambda^{-2 \nu}\nperp \otimes \nperp) \cdot   \mathrm{d}\mathbf{l}\equiv \mathrm{d}\mathbf{l} \cdot  \bar{a} \cdot \mathrm{d}\mathbf{l} \label{eqn:metric1}
\end{equation}
where $\nperp$ is orthogonal to $\n$. The sheet deforms to adopt this programmed metric, $\bar{a}$, in general becoming a curved surface such as Fig.~\ref{fig:theory}b. The sheet's small thickness $t$ also increases by $\lambda^{-\nu}$ upon activation. Curvature, $\kappa$, is penalised by a bending energy $\propto t^3 \kappa^2$, leading to residual bending stresses $\propto t \kappa$. In the $t \to 0$ limit, this bend cost is negligible compared to the cost of deviating from $\bar{a}$, which incurs stretch energy $\propto t$. Thus bend only enters as a `tie-breaker' between (thickness-independent) isometries of $\bar{a}$. 

Our shape programming task is to choose the director pattern $\n$ that will morph the annulus into our desired surface. We restrict our attention to axisymmetric director patterns, partly for simplicity, and partly because many use cases may actually be best suited by axisymmetric actuation. We define such patterns using the angle $\alpha(R)$ between $\n$ and the radial basis vector $\hat{\vec{R}}$:
\begin{equation*}
\n=\cos{(\alpha)} \,  \hat{\vec{R}}+\sin{(\alpha)} \, \hat{\vec{\theta}},\,\,\,\,\,\,\, \nperp=-\sin{(\alpha)} \,  \hat{\vec{R}}+\cos{(\alpha)} \, \hat{\vec{\theta}}.
\end{equation*}
The integral curves of this pattern are spirals, as shown in Fig.~\ref{fig:theory}a. Such spirals encode nematic bend and splay vector fields given by,
\begin{align}
&\vec{b}\equiv  (\nabla \times \n) \, \nperp = \left(\sin(\alpha)/R+\alpha'\cos{(\alpha)} \right) \, \nperp, \notag \\
&\vec{s}\equiv  (\nabla \cdot \n) \, \n = \left(\cos(\alpha)/R-\alpha'\sin{(\alpha)} \right) \, \n, \label{eqn:bendsplay}
\end{align}
and implicitly contain an $m=1$ topological defect at the origin, since the director winds by $2\pi$ on traversing any loop containing the hole.

The essential starting point for such metric design problems is Gauss's \emph{Theorema Egregium} \cite{gauss1828disquisitiones, o2014elementary}, which states that the Gauss curvature of the activated surface (computed as the product of the surface's principal curvatures, $K_A= \kappa_1 \kappa_2$) is an intrinsic geometric property and hence determined entirely by the metric. Thus the metric must encode the Gauss curvature of the target surface. A direct calculation using the LCE metric in a defect-free region yields \cite{aharoni2014geometry, mostajeran2015curvature,niv2018geometric, defective_nematogenesis}, in terms of gradients of bend and splay of the director field,
\begin{equation}
K_A= \half\left( \lambda^{2 \nu} - \lambda^{-2} \right)\nabla \cdot  \left(  \vec{b}+ \vec{s} \right),
\label{eqn:nontopGC}
\end{equation}
where all quantities on the right-hand side are evaluated in the reference state.  Working with bend and splay, which together offer a characterisation of a 2D director field, will turn out to offer particular insights into the genesis of the Gauss curvature.

We now set $K_A=0$, partly for simplicity again, but also with the intention of designing an `iris' actuator, for which the activated state is a flat annulus. Given axisymmetry, the divergence only acts on the radial component $(\vec{b} + \vec{s}) \cdot \hat{\vec{R}} =\cos(2 \alpha)/R-\alpha' \sin(2 \alpha)$ to give the equation
\begin{align*}
&\frac{1}{R}\frac{d}{d R}(R(\cos(2 \alpha)/R-\alpha' \sin(2 \alpha) ))=0,
\end{align*}
a result also obtained \cite{mostajeran2015curvature, kowalski2018curvature} from the general expression for $K_A$ in terms of $\alpha(r)$ and its gradients for spirals. It may immediately be integrated once with respect to $R$ to get
\begin{align*}
 \cos(2\alpha)/R-  \alpha' \sin(2 \alpha)=c_1/R.
\end{align*}
Finally, we solve for $\alpha$ yielding
\begin{equation}
\cos{(2 \alpha(R))}=c_1+c_2/R^2,\label{eq:pattern}
\end{equation}
where $c_1$ and $c_2$ are constants of integration. Previous authors \cite{mostajeran2015curvature, kowalski2018curvature} obtaining this result have then immediately set $c_2=0$ to avoid divergence as $R\to 0$. However, in an annular geometry $c_2$ is allowed, provided the pattern obeys $-1 \leq \cos(2 \alpha) \leq 1$, yielding the limiting radii
\begin{equation}
R_\pm^2=\frac{\pm c_2}{1\mp c_1},
\end{equation}
with radial and azimuthal alignment respectively for $\pm$.

However, setting $K_A=0$ within the LCE annulus does not guarantee a flat iris upon activation: we might have a cylinder or a truncated cone, neither of which can be flattened isometrically into the plane. Interestingly, this situation is fundamentally different to simply connected actuators, where, in accordance with Minding's theorem, setting $K_A=0$ everywhere guarantees an actuated surface that can be flattened isometrically into the plane. In the annular case, we thus require an additional intrinsic geometric property to select between these different Gauss-flat surfaces. 

Clarity is provided by geodesic curvature and the Gauss-Bonnet theorem. In general, the geodesic curvature, $k_g$, of a curve on a surface is computed by projecting its (3D) vector curvature into the tangent plane of the surface. For example, the dashed circle in Fig.~\ref{fig:theory}b has curvature $1/r$, but geodesic curvature $k_g$ of magnitude $\sin(\phi)/r$. Like Gauss curvature, geodesic curvature is intrinsic. Furthermore, geodesic curvature is connected to Gauss curvature by the Gauss-Bonnet theorem which, for any topologically-disk-like patch of surface, states that
\begin{equation}
 \Omega \equiv \int K \, \mathrm{d}A= 2 \pi -\oint k_g \, \mathrm{d}l  ,  \label{eq:GB}
\end{equation}
where the right-hand integral is around the boundary of the patch. Gauss-Bonnet thus allows us to compute the integrated curvature within a region of surface from the geodesic curvature on its boundary. If we have a surface containing a hole (e.g.~a truncated cone as in Fig.~\ref{fig:theory}b) we may imagine covering the hole with a smoothly connected patch. Gauss-Bonnet then reveals that this patch must contain integrated Gauss curvature
\begin{equation}
\Omega_{\circ}\equiv 2 \pi -\oint_{\mathrm{hole}} k_g \, \mathrm{d l},
\label{eqn:omega_hole}
\end{equation}
which is computed solely from the geodesic curvature of the hole's boundary, and is thus independent of nature of the patch. Furthermore, since $k_g$ is an intrinsic quantity, so is $\Omega_{\circ}$, which may be interpreted as the flux of Gauss curvature threading the hole. For example, the truncated cone in Fig.~\ref{fig:theory}b has  $\Omega_{\circ}=2\pi(1-\sin\phi)$ threading the hole, which is familiar as the integrated Gauss curvature of a cone-tip. The quantity $\Omega_{\circ}$ thus serves as a convenient additional intrinsic property in the shape programming of surfaces with holes, which is able to  distinguish cylinders ($\Omega_{\circ}=2 \pi$), truncated cones ($\Omega_{\circ}=2\pi(1-\sin\phi)$) and flat annuli $(\Omega_{\circ}=0)$.  

Returning to LCE shape-programming, the activated geodesic curvature $k_{gA}$ along any loop in the reference state may be computed from the LCE metric~\cite{defective_nematogenesis}, to reveal that 
\begin{align}
2\pi-\oint k_{gA} \, \mathrm{d}l =& \,\frac{1}{2} \left(\lambda^{1+\nu}-\lambda^{-1-\nu} \right)\oint  (\mathbf{b}+\mathbf{s})\cdot \vec{\hat{\nu}} \, \mathrm{d}l \notag \\&+ m\pi \left(1-\lambda^{1+\nu} \right) \left(1-\lambda^{-1-\nu} \right) ,\label{eq:Gaussgeneral}
\end{align}
where the boundary integral on the right is  conducted in the reference state,  $\vec{\hat{\nu}}$ is the reference state outward normal, and $m$ is the winding number of the director around the loop which, for a simply connected domain, would be the topological defect charge within. We note that our patterns have $ (\mathbf{b}+\mathbf{s})\cdot\mathbf{\hat{R}}=c_1/R$ and $m=1$. Thus, applying eq.~\ref{eq:Gaussgeneral} to the inner boundary of our annular reference domains, we compute that
\begin{equation}
\Omega_{\circ} = \pi\left(2-(1+c_1)\lambda^{-1-\nu}-(1-c_1)\lambda^{1+\nu}\right) ,
\end{equation}
revealing that the dimensionless constant $c_1$ alone controls $\Omega_{\circ}$, and selects between flat annuli, cylinders and truncated cones. This insight is immediately accessible from the form of $ (\mathbf{b}+\mathbf{s})\cdot\mathbf{\hat{R}}$ that determines the non-topological component of $\Omega_{\circ}$, highlighting the utility of working with bend and splay. In contrast, the constant $c_2$, which has dimensions of $\mathrm{length}^2$, simply determines the extent of the pattern in the reference domain, and hence the overall size of the actuated surface. In the following sections, we will explore the different modes of actuation achieved by programming $c_1$ and hence $\Omega_{\circ}$. We start with an extended discussion (and demonstration) of patterns for flat irises, cones and cylinders, and then map out the entire phase diagram produced by this set of patterns.

\section{Iris actuator: $\Omega_{\circ}=0$}

\begin{figure}[ht]
\centering
\includegraphics[width=\linewidth]{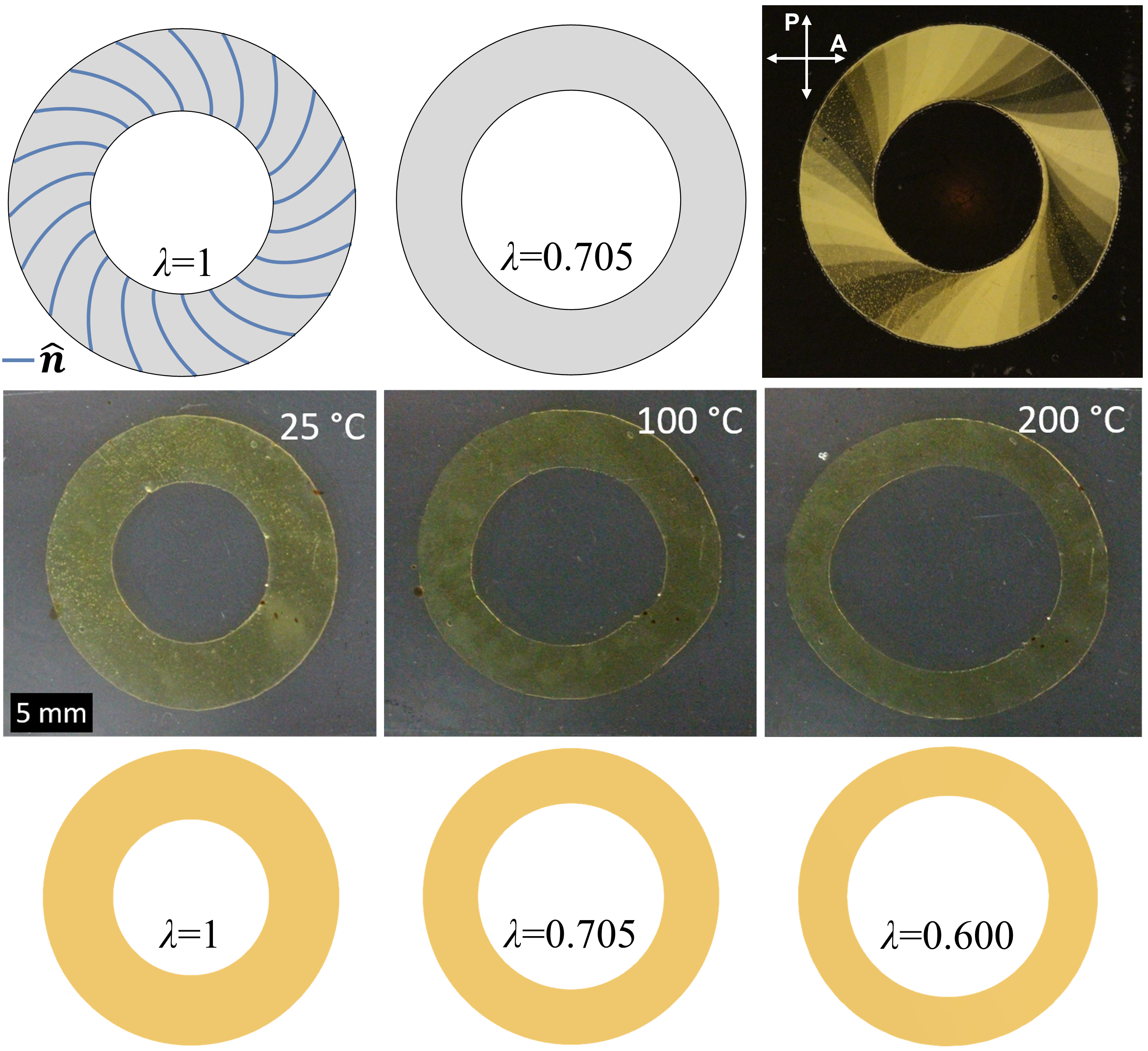}
  \caption{We design a spiral director pattern (top-left) that produces a flat anchored iris (top-middle) on actuation to $\lambda=0.705$. Top right shows a crossed-polars image of an LCE sample bearing this spiral pattern, with the dark and bright patterns associated with the director profile. Iris actuation is observed on heating the LCE from  from 25\,$^\circ$C to 200\,$^\circ$C (middle), with the design $\lambda$ attained near 100\,$^\circ$C. Numerical calculations (bottom) also confirm iris actuation, and show, surprisingly, that the iris remains flat at even more extreme actuation strains ($\lambda=0.6$, 200\,$^\circ$C) due to its non-zero bending stiffness. }
  \label{fig:iris}
\end{figure}

We start our exploration of annular actuators by demonstrating an iris: an annular LCE that activates to a flat annular state, but with a different inner radius. Such an iris could be used as an optical aperture, or to regulate flow down a pipe.

Previous work has demonstrated an LCE iris constructed from ``petals'' that bend away from the light-path to increase the aperture \cite{zeng2017self}. The beauty of this design was that the LCE was photo-active, and the bending was driven by the incident light, making the iris self-regulating. However, the resultant actuation is non-planar, leads to a non-circular aperture, and is mechanically weak as it depends on bend rather than stretch. A second approach has been to pattern an LCE annulus with a simple radial alignment \cite{schuhladen2014iris}, generating radial contraction on heating that mimics the pattern of muscular in a biological iris. However, radial patterns should actually buckle into anticones on actuation \cite{modes2011gaussian}, rather than remaining flat. This was avoided in \cite{schuhladen2014iris} by using an extremely thick sample, so that bending stiffness prevented out-of-plane deformation. Here, instead, we set $\Omega_{\circ}=0$, to target a truly flat annulus as the final state, which should remain flat even in the limit of a thin sheet, or at very high actuation strains. To achieve $\Omega_{\circ}=0$, we must choose:
\begin{equation} c_1 =  - (1- \lambda^{1+ \nu})/(1 + \lambda^{1+ \nu}).\label{eq:iris}
  \end{equation}
  This expression generates a unique $c_1$ for any actuation parameters $(\lambda,\,\,\nu)$. Furthermore, since any parameters generate  $-1 \le c_1 \le 1$, the resultant pattern is actually defined out to infinite radius, where it converges to a simple log-spiral (constant $\alpha$). However, if $c_2$ is included in the pattern, the pattern terminates at a finite inner radius, which is $R_+$ (radial director) for $c_2>0$ and $R_-$ (azimuthal director) for $c_2<0$. To design an iris with maximum dilation on actuation, we require radial alignment on the inner boundary, so we take $c_2>0$ and $R_i=R_+$.
  
Given the pattern extends out to infinity, we must also choose an outer boundary. To guide our choice, we note that, although elastomers are incompressible ($\nu=1/2$), LCE sheets reduce in area on actuation by a factor of $\sqrt{\lambda} = \lambda^{1-\nu}$, with overall volume conserved by associated thickening. In an LCE iris, although the hole radius dilates on actuation, circles at large radius must contract. There is thus a circle at an intermediate radius that is fixed on actuation, which can be found from area considerations:
\begin{equation}
R_{\mathrm{fix}}^2 - \lambda^{-2 \nu}R_i^2 = \lambda^{1-\nu}(R_{\mathrm{fix}}^2-R_i^2).
\label{eqn:Rfix}
\end{equation}
By choosing $R_{\mathrm{fix}}$ as the outer radius of the sample we attain an \textit{anchored iris actuator}, which could be fastened along its outer circumference in a frame or pipe, without suffering stretches in its final state (at the design $\lambda$) that could otherwise lead to buckling, damage etc.
  
To demonstrate this anchored iris actuation, we fabricate an annular LCE sheet with reference state radii ${R_i=5 \, \mathrm{mm}}$ and ${R_o=9.5 \, \mathrm{mm}}$. Given incompressibility, imposing $R_o=R_{\mathrm{fix}}$ yields $\lambda=0.705$, and hence (via eq.~\ref{eq:iris}) $c_1=-0.257$. Finally, setting $R_i=R_+$, fixes the value of $c_2$ and completes the specification of the pattern. 

Having specified the pattern, it is imprinted into the LCE using photo-patterning. To reduce the number of alignments, the continuous theoretical patterns were `binned' into 20 discrete angles before fabrication. The resulting iris was actuated by heating in a bath of silicon oil, and displayed very satisfactory dilation and anchoring, as shown in Fig.~\ref{fig:iris}.

We further verified this pattern by conducting numerics using our bespoke code MorphoShell~\cite{defective_nematogenesis} that, unlike our theory, accounts for both bend and stretch in a sheet with non-zero thickness. As seen in supplementary movie M1~\cite{duffy_annuli_movies_2021}, our numerics highlight that if its outer circumference is artificially pinned (cannot move) the iris remains planar and moves smoothly throughout its actuation without sudden instabilities, despite the stresses present before the design $\lambda$ is reached. If instead the boundary is completely free (supplementary movie M2), the iris passes through weakly conical shapes as it actuates, before regaining flatness at the design value of $\lambda=0.705$, consistent with attaining $\Omega_{\circ}=0$ only at the design $\lambda$. Similarly, as $\lambda$ falls below the design value, one would again expect non-flat shapes. However, as seen in Fig.~\ref{fig:iris}, both simulation and experiment indicate that, for realistic sheet thicknesses, energetic bend cost can suppress this transition, at least until $\lambda=0.6$. Additional simulations using `binned' director patterns (supplementary movie M3) confirm that the binning has little effect.

Our iris dilates upon heating (decreasing $\lambda$) and contracts upon cooling (increasing $\lambda$). One can instead create irises that contract on heating by setting $c_2<0$ so that the inner boundary has azimuthal director $R_i=R_-$. However, since the LCE must lose area during actuation, there is no fixed outer radius for anchoring in such systems. Similarly, it would also be impossible to create an anchored iris via isotropic swelling, unless swelling and shrinking co-occur in the same sample. However in nematic photo-glasses $\nu$ can range as high as 2, leading to areal growth on heating and enabling contractile anchored iris actuation, albeit with limited actuation strain. Within LCEs, oblate order parameters (where $\lambda > 1$ on heating) could offer another approach to such irises.

\section{Cones and Cylinders: $0 < \Omega_{\circ} \leq 2\pi$}

\begin{figure}[ht]
\centering
\includegraphics[width=\linewidth]{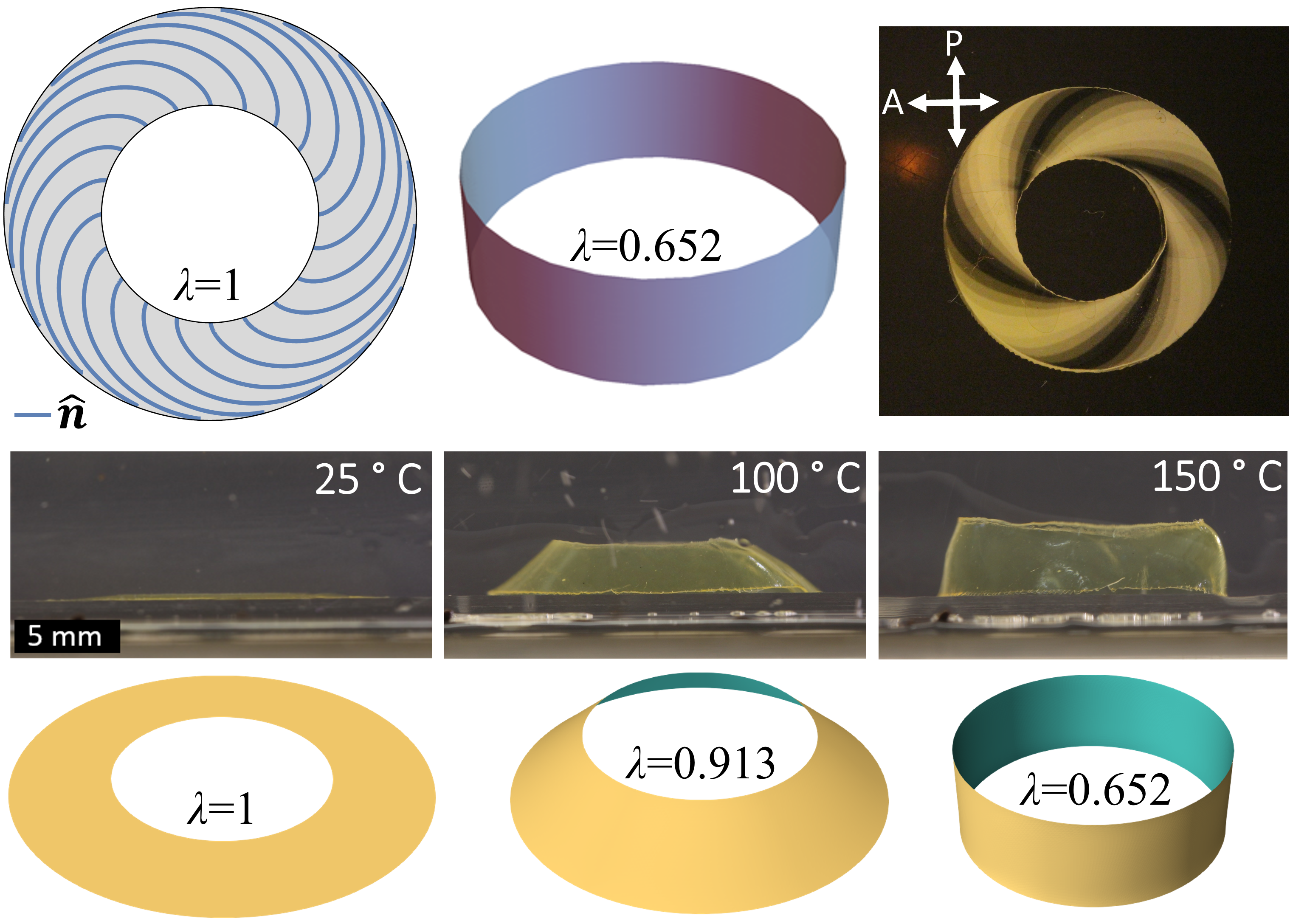}
  \caption{Director spiral (top-right) to morph an annulus into a cylinder (top-middle) at a design ${\lambda=0.652}$. Experiments (middle) and numerics (bottom) confirm the cylinder is attained, and also highlight the intermediate cone states en route. }
  \label{fig:cylinder}
\end{figure}

Any choice of $c_1$ other than that in eq.~\ref{eq:iris} will give $\Omega_{\circ} \ne 0$, producing shapes which are Gauss-flat but with integrated curvature threading the hole: truncated generalized cones. As discussed previously, a truncated cone has $\Omega_{\circ}=2 \pi(1-\sin\phi)$, with $\phi$ being the semi-angle, and $0<\Omega_{\circ}<2\pi$ parameterizing the range of cones between a flat annulus and a cylinder. Comparing this $\Omega_{\circ}$ to that of our patterns, we see that our patterns produce truncated cones with semi-angles given by
\begin{equation} 
\sin\phi =  \half \lambda ^{-1-\nu} \left(1+c_1+ (1-c_1)\lambda ^{2 +2 \nu}\right) , 
\label{eq:cone-angle}
\end{equation}
which is familiar \cite{modes2011gaussian} from log-spiral patterns with $\cos(2 \alpha)=c_1$. The underlying deformations, $(R,\theta)  \to (r(R), \, \theta+\Delta\theta(R), \, z(R))$, are non-linear functions of radius. Indeed, by comparing the resultant metric $a$ with the target metric $\bar{a}$, we may construct an exact isometry of $\bar{a}$:
\begin{align*}
r(R)^2 &= R^2\left(\lambda ^{-2 \nu } \cos ^2\alpha(R)+\lambda ^2 \sin ^2\alpha(R) \right),\\
z(R) &= \pm \cot(\phi)r(R),\\
\frac{\mathrm{d} \Delta \theta}{\mathrm{d} R} &= \frac{R}{2 r(R)^2}\left(\lambda^2-\lambda^{-2\nu}\right) \sin\left(2 \alpha(R)\right),
\end{align*}
which highlights the presence of twist between the inner and outer circumferences.

By varying $c_1$, we may attain a wide range of cone-angles. For simple log-spiral patterns, one is limited to $-1 \le c_1 \le 1$, and the steepest cone is given by $c_1=-1$ (a concentric circle pattern giving $\sin{\phi}=\lambda^{1+\nu}$). However, in annular domains, the addition of $c_2$ removes this limitation on $c_1$, so we can explore the full range of $\phi$. This enables sharper cones; indeed we can  even get cylinders  ($\Omega_{\circ}=2\pi$, $\sin\phi=0$) by setting in eq.~\ref{eq:cone-angle}
\begin{equation}
c_1 = -(1 + \lambda^{2+2\nu})/(1-\lambda^{2+2\nu}).\label{eq:cylinder}
\end{equation}
Such a cylinder pattern requires $c_1<-1$, so we must take $c_2>0$ yielding a pattern confined to the annular region, $R_+<R<R_-$, in which the director varies from radial to azimuthal, dilating the inner radius and contracting the outer one on heating. 

To demonstrate this cylindrical actuation,  we once again use the experimentally convenient domain ${R_i=5 \, \mathrm{mm}}$ and ${R_o=9.5 \, \mathrm{mm}}$. Solving $R_i=R_+$ and $R_o=R_-$ fixes $c_1=-1.766$ and $c_2$, and then eq.~\ref{eq:cylinder} reveals that a cylinder will be attained at $\lambda=0.652$, just within our experimentally accessible range. The resulting cylinder actuator is shown in Fig.~\ref{fig:cylinder}, alongside matching numerics (supplementary movies M4, M5). Actuation proceeds via sharpening cones as $\Omega_{\circ}$ rises from 0 to $2 \pi$. Here we use the full mathematical region of the cylinder pattern, $R_+<R<R_-$, however any annular sub-region could be selected to give a shorter cylinder. In particular, one can again find an invariant radius, $R_{\mathrm{fix}} = \lambda R_-$, that could be used to anchor the inner or outer boundary. This can be quickly found from the condition that, to produce a cylinder, every reference-state circle must attain the same activated-state radius, which then must equal $R_\mathrm{fix}$.

\section{Eversion, anticones, and the complete phase diagram}

\begin{figure*}[t]
\begin{centering}
\includegraphics[width=\textwidth]{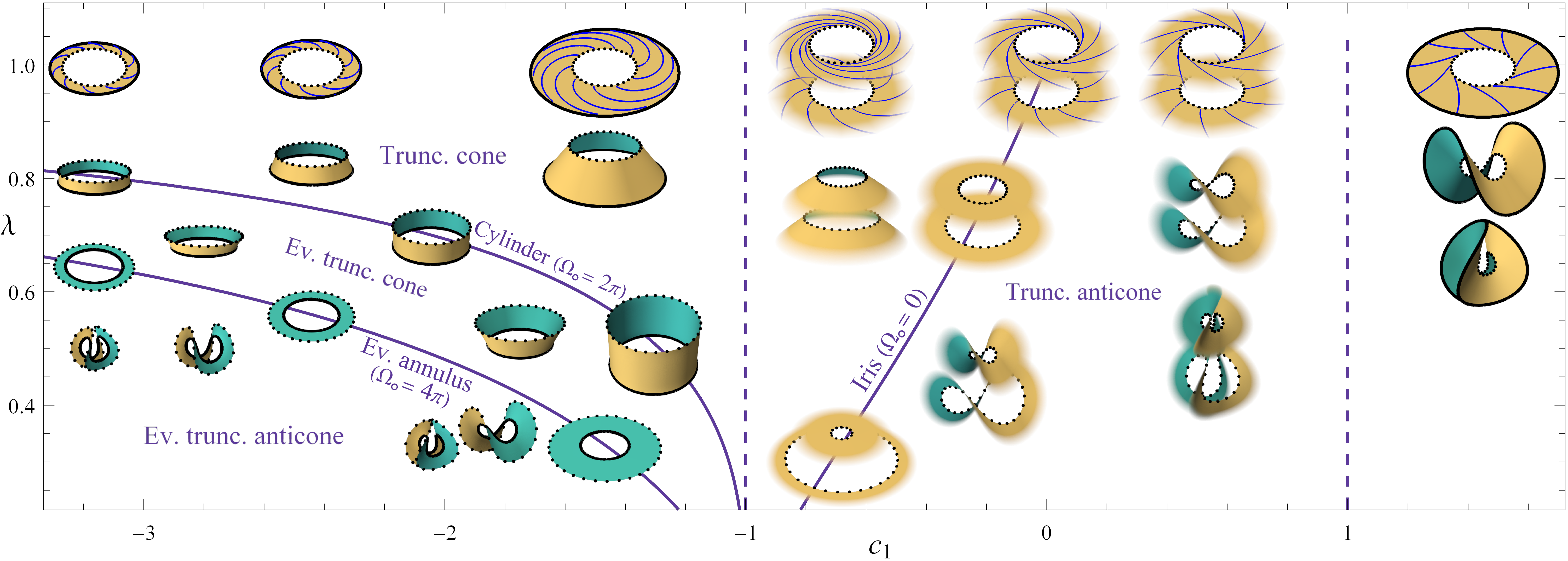}
\end{centering}
\caption{Phase diagram for annular LCE actuators in $c_1 - \lambda$ space (for $\nu=1/2$). The top row ($\lambda=1$) shows the un-actuated director patterns given by different values of $c_1$. For $-1<c_1<1$ the pattern converges to a simple log spiral at large radius, but terminates at a finite inner radius with an azimuthal or radial director depending on the sign of $c_2$. Both possibilities are shown in this region. For $c_1>1$ one must take $c_2<0$, and the pattern is defined in a finite annulus with azimuthal director at the inner boundary, while for $c_1<-1$ one must take $c_2>0$, and the pattern is defined on a finite annulus with radial director at the inner boundary. The remainder of the diagram shows simulations of the actuated shapes achieved when $\lambda \ne 1$, taking the full extent of the pattern in the finite cases. Dotted and bold lines differentiate the initially-inner and initially-outer boundaries, while `fade-out' regions indicate surfaces extending to infinity. The diagram reveals regions of truncated cones, truncated anticones, everted truncated cones and everted truncated anticones, which are classified by intervals of $\Omega_{\circ}$, and are separated by by lines of irises ($\Omega_{\circ}=0$), cylinders ($\Omega_{\circ}=2 \pi$), and everted annuli ($\Omega_{\circ}=4 \pi$). The iris, cylinder, and everted annulus lines are given by theory, and  meet at $\lambda=0$.}
\label{fig_phase_diagram} 
\end{figure*}

Finally we classify the full set of actuators available via eq.~\ref{eq:pattern}. Any value of $\Omega_{\circ}$ can be achieved with a suitable choice of $c_1$. However, for $c_1>1$ we must take $c_2<0$, giving an annular region $R_-<R<R_+$ with azimuthal (radial) director at the inner (outer) boundary. Conversely, for $c_1<-1$  we must take $c_2>0$, giving an annular region $R_+<R<R_-$ with radial (azimuthal) director at the inner (outer) boundary. Between these limits, $-1\le c_1 \le 1$, the patterns converge to a  log-spiral with constant angle $\alpha(\infty)=\half\cos^{-1}(c_1)$  as $R\to\infty$. However, the $c_2$ term will still diverge towards the origin until the director is purely radial (+ve $c_2$) or azimuthal (-ve $c_2$), requiring $R_\pm<R<\infty$ respectively. In every case, the magnitude of $c_2$ (with units of area) simply sets the overall scale of the pattern without affecting its proportions, or the shape of the resultant surface. Without loss of generality we can thus display the full family of patterns  in eq.~\ref{eq:pattern} parameterized only by $c_1$ and the sign of $c_2$, as shown along the top of Fig.\ \ref{fig_phase_diagram}, where $\lambda = 1$ (the reference state).

Upon heating, $\lambda$ will diminish below unity, and each annulus will morph into an increasingly extreme surface, generating a $c_1-\lambda$ phase diagram that captures the full set of actuated shapes. As seen in  Fig.~\ref{fig_phase_diagram}, the various regions of the phase diagram can be straightforwardly identified as intervals of $\Omega_{\circ}$. We already have a cone region $0<\Omega_{\circ}<2\pi$. For $\Omega_{\circ}<0$ we have surfaces with an angular surplus rather than an angular deficit, which buckle into ruff-like (truncated) anticones. Moving in the other direction, the cone region terminates at $\Omega_{\circ}=2\pi$ with a line of cylinders. However, there is nothing preventing $c_1$ being chosen to make $\Omega_{\circ}$ still larger. For $2\pi<\Omega_{\circ}<4\pi$ we will have a region of everted truncated cones ($-1<\sin\phi<0$) in which the original outer boundary becomes the actuated inner boundary. Everted cones terminate at $\Omega_{\circ}=4\pi$ (i.e.~$\sin \phi=-1$) with a line of everted flat annuli,
\begin{equation}
c_1 = -(1 + \lambda ^{1+\nu})/(1 - \lambda ^{1+\nu}).
\end{equation}
Finally, for $\Omega_{\circ} > 4\pi$ we have the most extreme mode of actuation: everted anticones.

During heating, an LCE annulus will descend a vertical (constant $c_1$) line of diminishing $\lambda$ on the phase diagram. We again use MorphoShell to calculate a spectrum of such actuation pathways (supplementary movies M6-M10), and graphically populate the phase diagram for $\nu=1/2$. Interestingly, a given annulus may move through several regions during actuation, and thus show several qualitatively different shapes en route to the target. For $-1<c_1<0$ one has flat $\to$ cone $\to$ iris $\to$ anticone (e.g.~M8), while  $c_1<-1$  gives flat $\to$ cone $\to$ cylinder $\to$ everted cone $\to$ everted annulus $\to$ everted anticone (e.g.~M6). Such behaviour hints at complex nematic mechanisms with multiple useful shapes realized at different temperatures. For example, both paths start flat, rise out of the plane, then regain flatness, yielding a non-monotonic height change that could serve as a frequency doubler for a cyclic stimulus.

The phase diagram can be extended to $\lambda > 1$, and/or replotted for different values of the opto-thermal Poisson ratio $\nu$, including $\nu<0$ as can be realised by materials that swell or shrink anisotropically. However, no qualitatively new activated shapes emerge, which can be understood from eq.~\ref{eqn:metric1} as follows: First~(\dag), observe that rotating $\n$ and $\nperp$ by $\pi/2$ whilst simultaneously interchanging ${\lambda \leftrightarrow \lambda^{-\nu}}$ leaves $\bar{a}$ unchanged. Second~(\ddag), observe that $\bar{a}$ can be written in the form 
\begin{equation}
\bar{a} = \Lambda^2 (\tilde{\lambda}^2 \, \n \otimes \n + \tilde{\lambda}^{-1}  \, \nperp \otimes \nperp) ,
\end{equation}
corresponding to an anisotropic deformation of ${\tilde{\lambda} = \lambda^{2(1+\nu)/3}}$ along $\n$ with an opto-thermal Poisson ratio of 1/2, followed by an isotropic lineal scaling ${\Lambda=\lambda^{(1-2\nu)/3}}$. By applying either~(\dag), or~(\dag) followed by~(\ddag), a pattern with any ${\lambda, \, \nu}$ can be be seen to result in the same activated shape (up to an overall scale factor) as a pattern with ${\lambda <1, \,  \nu=1/2}$, the parameter regime explored in Fig.\ref{fig_phase_diagram}. 

If ${\nu > -1}$, then for ${\lambda > 1}$ the cylinder and everted annulus lines have ${c_1>1}$, and asymptote to ${\lambda = 1}$ as ${c_1 \to \infty}$ and ${c_1=1}$ as ${\lambda \to \infty}$, while the iris line continues smoothly on from its ${\lambda<1}$ portion, also asymptoting to ${c_1=1}$ as ${\lambda \to \infty}$.
% Useful math accents for future reference:
%https://tex.stackexchange.com/questions/177000/math-mode-accents

Returning to the question of anchoring, if $\nu < 0$ then either both $\lambda$ and $\lambda^{-\nu}$ are $>1$, or both are $<1$. Clearly then no choice of director can match the azimuthal distortion to that of an inert support, i.e.~ one cannot achieve zero azimuthal distortion, and stress-free anchoring is impossible. However, if $\nu > 0$ and
\begin{equation}
\mathrm{sgn}(c_2) \left( c_1 - \frac{\lambda^2 + \lambda ^{-2\nu} - 2} {\lambda^2 - \lambda^{-2\nu}} \right) < 0 ,
\label{eqn:RfixExistenceCond}
\end{equation}
one can again find an $R_\mathrm{fix}$ at which there is zero azimuthal distortion. As long as $\nu>0$, the condition~(\ref{eqn:RfixExistenceCond}) is satisfied for any ${ |c_1|>1 }$, and for exactly one of the two possible signs of $c_2$ at each ${ |c_1| < 1 }$. %(which sign of $c_2$ depends on $\lambda$ and $\nu$). 
When actuation is axisymmetric ($0 \leq \Omega_\circ \leq 4\pi$), $R_\mathrm{fix}$ could be used to anchor the actuator as discussed previously.

For our non-axisymmetric actuators the situation is more subtle; such anchoring can only be stress-free if a suitably smooth embedding of $\bar{a}$ exists in which the `anchored' curve is a circle. However, no such embedding exists, as some toying with a paper truncated anticone will quickly demonstrate. To prove this, consider any closed reference-state curve that encloses the origin (e.g.~a circle at $R_\mathrm{fix}$), and suppose that upon activation this curve forms a circle. Observe that if the activated-state circle has radius $\Gamma_A$, then its curvature vector has magnitude $1/ \Gamma_A$.

Given that the geodesic curvature $k_{gA}$ is found by projecting the curvature vector into the tangent plane, it has magnitude ${|k_{gA}| \leq 1/ \Gamma_A}$. Then, applying Gauss-Bonnet (in the manner of eq.~\ref{eqn:omega_hole}) to the material within any such circle, we find ${\lvert 2 \pi - \Omega_\circ \rvert = \lvert \oint k_{gA} \mathrm{d}l_A \rvert \leq 2 \pi}$. Thus, for ${\lvert 2 \pi - \Omega_\circ \rvert > 2 \pi}$ the activated surfaces indeed \textit{cannot} accommodate such circles without stretching (deviating from $\bar{a}$).

\section{Discussion and Conclusions}
We have explored a new category of nematic actuators with an annular geometry and found that this change in topology introduces qualitatively different and more extreme modes of actuation. For example, although we have focused on patterns that do not encode Gauss curvature in the LCE region, many of the resultant surfaces are necessarily curved owing to a concentration of Gauss curvature within the hole. Furthermore, this `ghost' Gauss curvature is unconstrained by material considerations and can take any value, allowing a full spectrum of  (truncated) cones, irises, cylinders, anticones, and even their everted counterparts. In all cases the actuation is underpinned by large-strain \textit{metric mechanics}, guaranteeing mechanically strong actuation.

Although our design approach is metric-based, our numerics account for both stretch and bend, essentially providing bend-minimising isometries of the metric $\bar{a}$, as would be adopted by a thin sheet. Reassuringly, both here and in ref.\cite{kowalski2018curvature}, the axisymmetric theoretical isometries (e.g.~irises, truncated cones, cylinders) appear to be the bend minimisers, apparently validating a design approach in which the metric $\bar{a}$ is matched to a target surface. However, in general there are many isometries of a given $\bar{a}$, and in cases with less symmetry there is little reason to think that the bend-minimiser will be close to the target surface. Furthermore, the presence of many isometries means that even Gauss-curved surfaces can deform via energetically cheap pure-bending modes, and care must be taken to avoid such modes undermining strong actuation.

Looking ahead, it is natural to wonder about geometries with multiple holes, and patterns that lack axisymmetry or encode distributed Gauss curvature. Can one solve inverse problems in domains with holes? However, these lines of thought seem more likely to produce increasing complexity rather than qualitatively new modes of actuation. More promising perhaps is to consider the programming of initially curved surfaces, including closed spherical surfaces and even non-orientable M{\"o}bius strips. Such programmed shells would again have a genuinely different topology, inviting new modes of actuation, and could be fabricated by newly developed 3D-printing techniques.

\section{Materials and Methods}% This could be included in a supplement or not 
LCE films were prepared between glass plates separated with a 50$\, \mu$m spacer where the interior surfaces of the plates are coated with a photoalignable dye. First, the glass plates were cleaned by sequential sonication in Alconox-water solution, acetone, isopropanol, and deionized water. The glass slides were then exposed to oxygen plasma reactive ion etching for 1 min at 100$\,$mTorr pressure and 50$\,$mW power. The photoalignable dye solution, 1$\,$wt.\%  brilliant yellow in dimethylformamide, was spin-coated onto the glass plates at 750$\,$rpm with an acceleration of 1500$\,$rpm/s for 10$\,$s and then at 1500$\,$rpm with an acceleration of 1500$\,$rpm/s for 30$\,$s. Two glass plates are then adhered using a cyanoacrylate adhesive. The dye was then locally oriented on both plates by exposure to linearly polarized broadband visible light using a modified projector (Vivitek D912HD) such that the resolution of the exposure is 30$\, \mu$m. To create patterned alignment, the dye on regions of the plates were sequentially exposed using different polarization angles. A monomer solution was prepared using a liquid crystal monomer, 1,4-bis-[4-(6-acryloyloxyhexyloxy)-benzoyloxy]-2-methylbenzene (RM82, Wilshire Chemicals), a chain extender, n-butylamine (Sigma Aldrich), and a photoinitiator, Irgacure I-369 (BASF) by heating to 90$\, ^\circ$C and then vortex mixing the fluid. The molar ratio of RM82 to n-butylamine was 1.1:1. The photoinitiator was added at 1.5$\,$wt.\% of the monomer mixture. The solution was then filled by capillary action between the glass plates at 75$\, ^\circ$C. The sample was then stored at 65$\, ^\circ$C for 12$\,$hr for chain extension to occcur. The sample was then exposed to UV light at room temperature (OmniCure$^\text{®}$ LX400+, 250$\,$mW/cm2, 365$\,$nm) to crosslink the LCE. The total time of UV exposure was 5$\,$min and the sample was flipped at 2.5$\,$min of exposure. After crosslinking, one glass plate was removed, and the sample was cut from the surrounding regions using a CO$_2$ laser cutter (Universal Laser Systems ILS9.150D). The alignment was confirmed using optical imaging of transmitted light with the sample between crossed polarizers. To measure shape change, samples were immersed in silicone oil heated to the appropriate temperature and then imaged with a DSLR Canon camera.\vspace{0.0cm}\\

	{\centering \bf Data Availablity 
	\par
	}
The data that support the findings of this study are available from the corresponding author upon reasonable request.

\begin{acknowledgments}
D.D. was supported by the EPSRC Centre for Doctoral Training in Computational Methods for Materials Science [grant no. EP/L015552/1]. M.W. was supported by the EPSRC [grant number EP/P034616/1]. J.S.B. was supported by a UKRI “future leaders fellowship” [grant number MR/S017186/1].\nl
This material is partially based upon work supported by the National Science Foundation under Grant No. DMR~2041671.
\end{acknowledgments}
  
\bibliography{references.bib}

\begin{thebibliography}{39}
\bibitem [{duf()}]{duffy_annuli_movies_2021}%
  \BibitemOpen
  \href@noop {} {}\bibinfo {note} {Movies of our MorphoShell simulations are
  provided as ancillary files M1-M10.}\BibitemShut {Stop}%
\end{thebibliography}

\end{document}